\documentclass[aps,prl,twocolumn,showpacs,superscriptaddress,preprintnumbersfleqn,floatfix]{revtex4}

\usepackage{graphicx}
\usepackage{dcolumn}
\usepackage{bm}
\usepackage{amsmath, amssymb}

\begin{document}

\title{Microscopic parameters of the van der Waals CrSBr antiferromagnet from microwave absorption experiments}

\author{C. W. Cho}
\affiliation{Laboratoire National des Champs Magn\'etiques Intenses, CNRS, LNCMI, Universit\'e Grenoble Alpes, Univ Toulouse 3, INSA Toulouse, EMFL, F-38042 Grenoble, France}
\author{A. Pawbake}
\affiliation{Laboratoire National des Champs Magn\'etiques Intenses, CNRS, LNCMI, Universit\'e Grenoble Alpes, Univ Toulouse 3, INSA Toulouse, EMFL, F-38042 Grenoble, France}
\author{N. Aubergier}
\affiliation{Laboratoire National des Champs Magn\'etiques Intenses, CNRS, LNCMI, Universit\'e Grenoble Alpes, Univ Toulouse 3, INSA Toulouse, EMFL, F-38042 Grenoble, France}
\affiliation{Universit\'e Grenoble Alpes, CEA, Grenoble INP, IRIG, Pheliqs, 38000 Grenoble, France}
\author{A. L. Barra}
\affiliation{Laboratoire National des Champs Magn\'etiques Intenses, CNRS, LNCMI, Universit\'e Grenoble Alpes, Univ Toulouse 3, INSA Toulouse, EMFL, F-38042 Grenoble, France}
\author{K. Mosina}
\affiliation{Department of Inorganic Chemistry, University of Chemistry and
Technology Prague, Technick\'a 5, 166 28 Prague 6, Czech Republic}
\author{Z. Sofer}
\affiliation{Department of Inorganic Chemistry, University of Chemistry and
Technology Prague, Technick\'a 5, 166 28 Prague 6, Czech Republic}
\author{M. E. Zhitomirsky}
\affiliation{Universit\'e Grenoble Alpes, CEA, Grenoble INP, IRIG, Pheliqs, 38000 Grenoble, France}
\affiliation{Institut Laue-Langevin,  F-38042 Grenoble Cedex 9, France}
\author{C. Faugeras}
\affiliation{Laboratoire National des Champs Magn\'etiques Intenses, CNRS, LNCMI, Universit\'e Grenoble Alpes, Univ Toulouse 3, INSA Toulouse, EMFL, F-38042 Grenoble, France}
\author{B. A. Piot}
\affiliation{Laboratoire National des Champs Magn\'etiques Intenses, CNRS, LNCMI, Universit\'e Grenoble Alpes, Univ Toulouse 3, INSA Toulouse, EMFL, F-38042 Grenoble, France}

\date{\today}

\begin{abstract}
Microwave absorption experiments employing a phase-sensitive external resistive detection are performed
for a topical van der Waals antiferromagnet CrSBr. The field dependence of two
resonance modes is measured in an applied field parallel to the three principal crystallographic directions,
revealing anisotropies and magnetic transitions in this material. To account for the observed results, we formulate a microscopic spin model with a bi-axial single-ion anisotropy and inter-plane exchange.
Theoretical calculations give an excellent description of full magnon spectra enabling us to precisely
determine microscopic interaction parameters for CrSBr.
\end{abstract}
\maketitle

The observation of ferromagnetism in two dimensions \cite{Huang2017,Gong2017} and the discovery of magnetically and electrically tunable van der Waals-bond few-layers compounds
(see Ref.~\onlinecite{Wang2022} for a review) have boosted the interests in the so-called van der Waals (vdW) magnets new class of materials. This has stimulated the search for new layered magnetic systems, as well as expanded the exploration of partially studied ones, even in their bulk forms. Among vdW magnets, layered antiferromagnets such as CrI$_{3}$, CrCl$_{3}$, CrSBr have the particularity of hosting a low temperature anti-ferromagnetic order associated with an alternating magnetic moment orientation between adjacent layers. At variance with typical (covalently bond) antiferromagnets, the weak interlayer exchange coupling in these vdW systems allows for the use of moderate magnetic fields to manipulate the N\'eel vector.  In CrSBr, there exists a significant additional bi-axial anisotropy which further defines a preferred in-plane magnetic moment direction (along the so-called easy axis). CrSBr has the further assets to be an air-stable semiconductor, the thickness of which can be adjusted down to the monolayer with a demonstrated coupling between magnetic order and charge transport \cite{Telford2022}. This makes it a promising system for integration in spin-based electronic devices \cite{Wang2022}, which motivates the need for an accurate description of its internal microscopic parameters.

In this article, we present a microwave absorption study of the low energy magnon excitations in bulk CrSBr which employs a phase-sensitive external resistive detection technique. Strong resonant signals are observed in a large frequency/magnetic field phase space, for frequencies continuously tuned up to $60$~GHz and magnetic fields up to $10$~T. The magnetic field dispersion of the two observed magnon modes strongly depends on the field orientation and can be understood by ``in-plane'' anisotropies and exchange interactions. The study of the magnetic field dependence
along each of the three crystallographic axes and up to well-above the saturations fields, together with a microscopic model based on single ion anisotropy and exchange, enable us to precisely extract
the microscopic magnetic parameters of this promising system.

\begin{figure*}[t]
\begin{center}
\includegraphics[width=18cm]{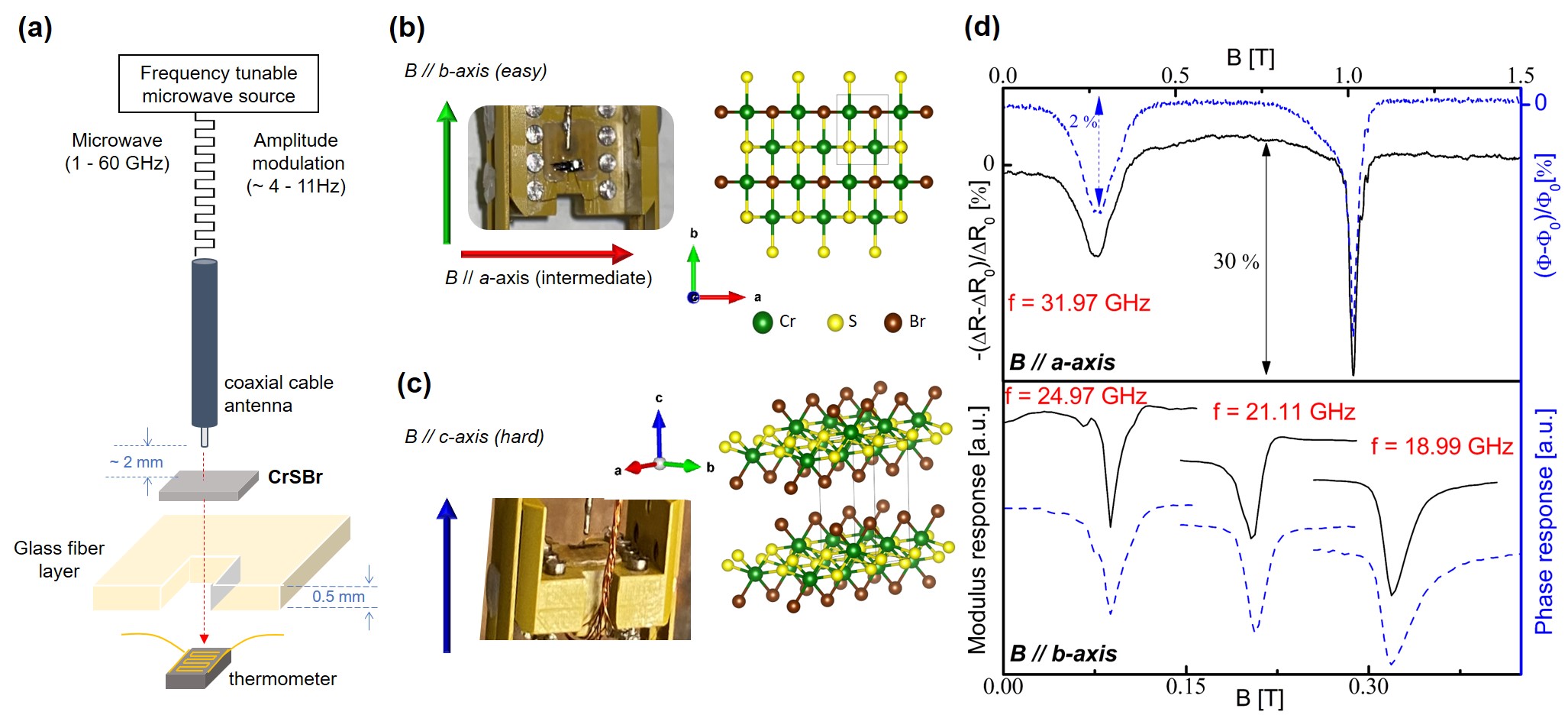}
\end{center}
\caption{(color online) (a) Schematic representation of the experimental microwave absorption setup. (b)``In-plane'' magnetic field $B$ configurations, with $B$ either along the $b$ (easy) axis or the $a$ (intermediate) axis, and the CrSBr crystal structure. (c)``Out-of-plane '' magnetic field $B$ configuration, with $B$ along the $c$ (hard) axis, and the CrSBr crystal structure. Photos in (b) and (c) show the actual sample environment.(d)
Representative raw data of magnetic field sweep-induced resonances. Upper panel: resonances observed for different magnons modes in the $B \parallel $ a-axis configuration at a \textit{single} microwave frequency. The detection can be made either through the relative change of the photo response amplitude $\Delta R$ with respect to its ``off-resonant'' value $\Delta R_{0}$ (solid black lines), or the relative change of the phase $\phi$ with respect to its ``off-resonant'' value $\phi_{0}$ (dashed blue lines). Lower panel: a single magnon mode resonance followed versus magnetic field for different microwave frequencies.
 }\label{Fig1}
\end{figure*}

The lattice of CrSBr consists of vdW layers made of two buckled planes of CrS terminated by Br atoms, which are stacked along the c-axis in an orthorhombic structure with Pmmn (D2h) space group, see figures~\ref{Fig1}(b) and ~\ref{Fig1}(c). The biaxial magnetic anisotropy is characterized by an easy axis along the crystallographic $b$ axis, an intermediate axis along $a$, and a hard axis along $c$. The bulk CrSBr crystals studied here were grown by chemical vapor transport \cite{SM.sample} and mechanically exfoliated in thin flakes samples of typically a few micrometers thick.

Magneto-absorption experiments were conducted in the GHz regime by monitoring the bolometric response of a resistive sensor placed below the sample, as can be seen in a schematic representation of our setup in Fig.~\ref{Fig1}(a).
Both sit in a $^4$He exchange gas environment inside a closed socket, inserted in a variable temperature insert (VTI) fitted into a 16 T superconducting magnet. To enhance the measurement sensitivity and minimize thermal drifts, a low frequency modulation of the microwave was employed, and the thermometer response was measured with a standard low frequency AC lock-in technique. In the case of modulated microwave, this photo-response can be studied both in terms of amplitude ($\Delta R$ and its modulus $|\Delta R|$) or phase ($\phi$) of the AC signal. The thermometer AC voltage response is generally anti-phased ($\Delta R<0$) with respect to the microwave modulation reference, due to the insulator-like behaviour of the sensor (R decreases as the temperature increases). In practice, there is a finite thermalisation time for the thermometer, which makes the actual phase of the signal different from $\pi$. As a microwave absorption takes place in the sample, an increase in the modulus $|\Delta R|$ of the photo response signal reveals a lower effective temperature of the thermometer (since the first derivative of $R(T)$ increases at lower $T$), consistent with the fact that a fraction of the incoming power has been absorbed by the sample. The phase of the signal also undergoes a change during the absorption process, which we interpret as a change in thermalisation times, again due to the sample absorption.

Figure~\ref{Fig1}(d) shows typical examples of the detection of different sample microwave absorption through the thermometer AC photo-resistance. In the upper panel, we show resonances observed for
a single (fixed) microwave frequency, where the two resonances corresponds to different magnon modes. The lower panel illustrates the magnetic field shift of a given magnon mode resonance as the frequency is changed. As we can see, in this technique, microwave absorption can generally be identified by a clear single peak function-like response. This is particularly true for the phase detection which, while showing smaller relative variations than the modulus, has the advantage of exhibiting a ``flat''  non-resonant background almost unaffected when sweeping the magnetic field. The linewidth of the resonances also gives information about the local curvature of the magnon mode (see for example the different linewidths for the two resonances reported in the upper panel), further demonstrating the high sensitivity of the measurement. In the remaining of this paper, we will therefore base our analysis on the \textit{phase response} which provides higher resolution data.

\begin{figure*}[]
\begin{center}
\includegraphics[width=18cm]{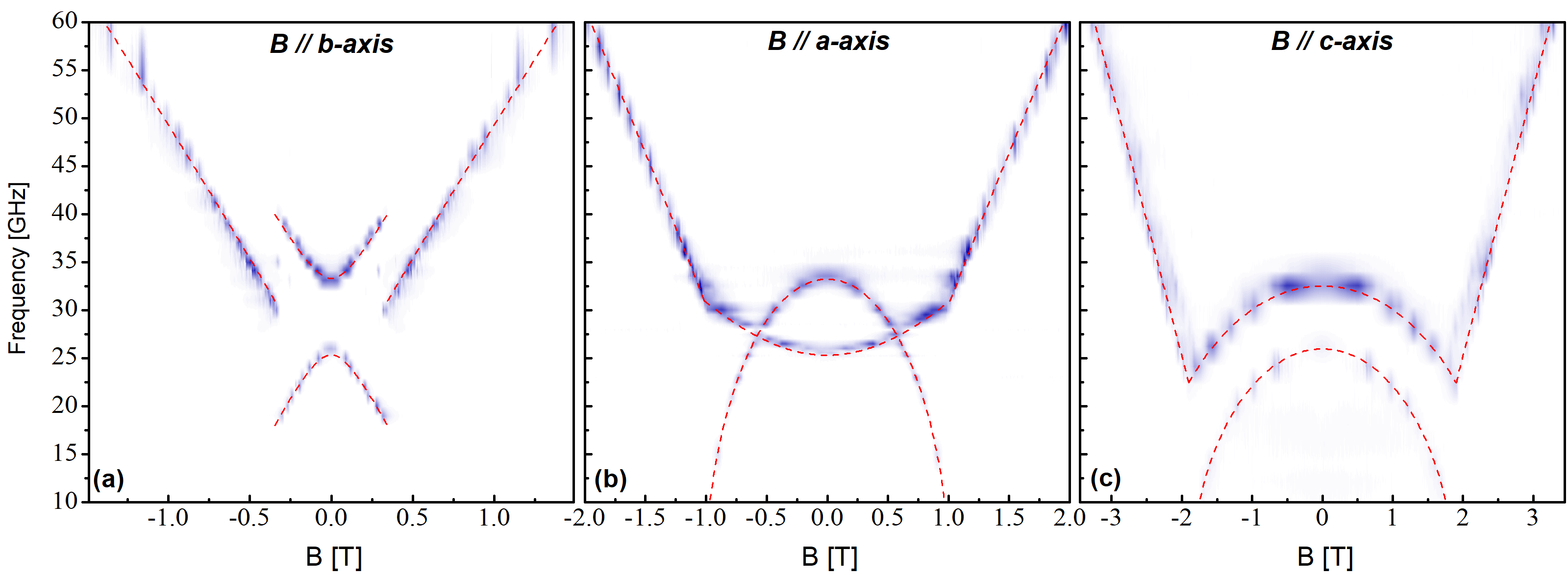}
\end{center}
\caption{(color online) Absorption spectra for a magnetic field $B$ along the b (left panel (a)), a (center panel (b)) and c (right panel (c)) axis. Color plots of the phase variation of the thermometer photo-resistance as a function of microwave frequency and magnetic field $B$. The phase difference with respect to its median value (which corresponds to the ``off-resonance'' value) is color-coded from zero (white) up to an upper value of 8\% (dark blue). Fits (red dashed-lines) obtained with our microscopic model (see text for details).}\label{fig2bis}
\end{figure*}

In the absence of magnetic field, two main absorptions can be observed at energies of about $26.05 \pm 0.01 $ GHz  and  $33.37 \pm 0.01$ GHz at our lowest temperature $T=2.2$ K \cite{SM.zeroB}, corresponding to the zero-field magnon modes recently identified in Refs.~\onlinecite{Cham2022,Bae2022}. In figure~\ref{fig2bis}(a), we present the magnetic-field dependence of these modes when an external magnetic field is applied parallel to the b (easy)-axis, along which magnetic moments are naturally oriented, at an effective temperature of $T = 5.4$ K.

The displayed color map is a linear extrapolation of the ``frequency/magnetic field'' dependence of the variations of the phase of the thermometer photo-resistance, obtained by sweeping the magnetic field for a fixed microwave frequency. The high energy $B=0$ mode disperses positively as $B$ increases, while the lower energy $B=0$ disperses negatively, both in a similar fashion, consistent with very recently reported cavity absorption measurements \cite{Cham2022}. Above a certain magnetic field ($B\sim 0.35$~T at $T=5.4$~K), an abrupt change in the spectrum is observed: the two modes suddenly disappear and a new single mode emerge from a frequency of about 29.8 GHz, and disperse approximatively linearly up to the highest studied frequencies (60 GHz). This change is concomitant with the magnetic transition from an antiferromagnetic ground state to a ferromagnetic one, reported in early magnetic measurements \cite{Goeser1990}, and more recently by transport measurements \cite{Telford2022}. This transition, which was also recently found to have consequences on the optical properties~\cite{Wilson2021,Cenker2022}, strongly modifies the magnon spectrum due to a change in the magnetic ground state (magnetic moments antiparallel to the applied $B$ abruptly switch to the parallel orientation).
We note that, for temperatures $T <$ 20 K at least, this transition is abrupt with no intermediate behaviour: the resonant frequency evolves monotonously with $B$ as soon as the switching field is reached, and disperses approximatively linearly up to the highest studied magnetic field/frequencies. This absence of an intermediate spin-flop phase is consistent with early magnetic susceptibility measurements \cite{Goeser1990}, and will be discussed theoretically below.

We now turn to the magnetic field dependence of the magnon modes when the magnetic field is applied along the a-(intermediate) axis. Figure~\ref{fig2bis}(b) shows the absorption spectrum in the frequency/magnetic field space in this configuration at an effective temperature of $T = 5.4$~K. The displayed color map is again a linear extrapolation of the magnetic field dependence of the thermometer \textit{phase} signal, taken for a fixed microwave frequency. In this configuration, magnetic moments progressively cant from their initial preferred (b-axis) orientation towards the a-axis alignment. A magnetic-field-induced crossing of the two branches, qualitatively similar to what is observed in the easy-plane CrCl$_{3}$ anti-ferromagnetic insulator \cite{MacNeill2019}, occurs at about 0.56 T. The presence of a bi-axial anisotropy nevertheless makes the $B=0$ situation quite different here, with, as we described earlier, the existence of a finite energy lower branch at $B=0$ (the latter tending to zero energy in CrCl$_{3}$). As the magnetic field is further increased after the crossing, the (initially) upper energy magnon mode vanishes to zero energy and the (initially) lower energy mode dispersion suddenly changes dependence. This change, occurring at $B \sim1$T, corresponds to the full saturation of magnetic moments along the a-axis. Above this saturation field, the frequency of the (initially) lower branch evolves approximatively linearly up to the highest studied magnetic field/frequencies.

\begin{figure}[]
\begin{center}
\includegraphics[width= 8cm]{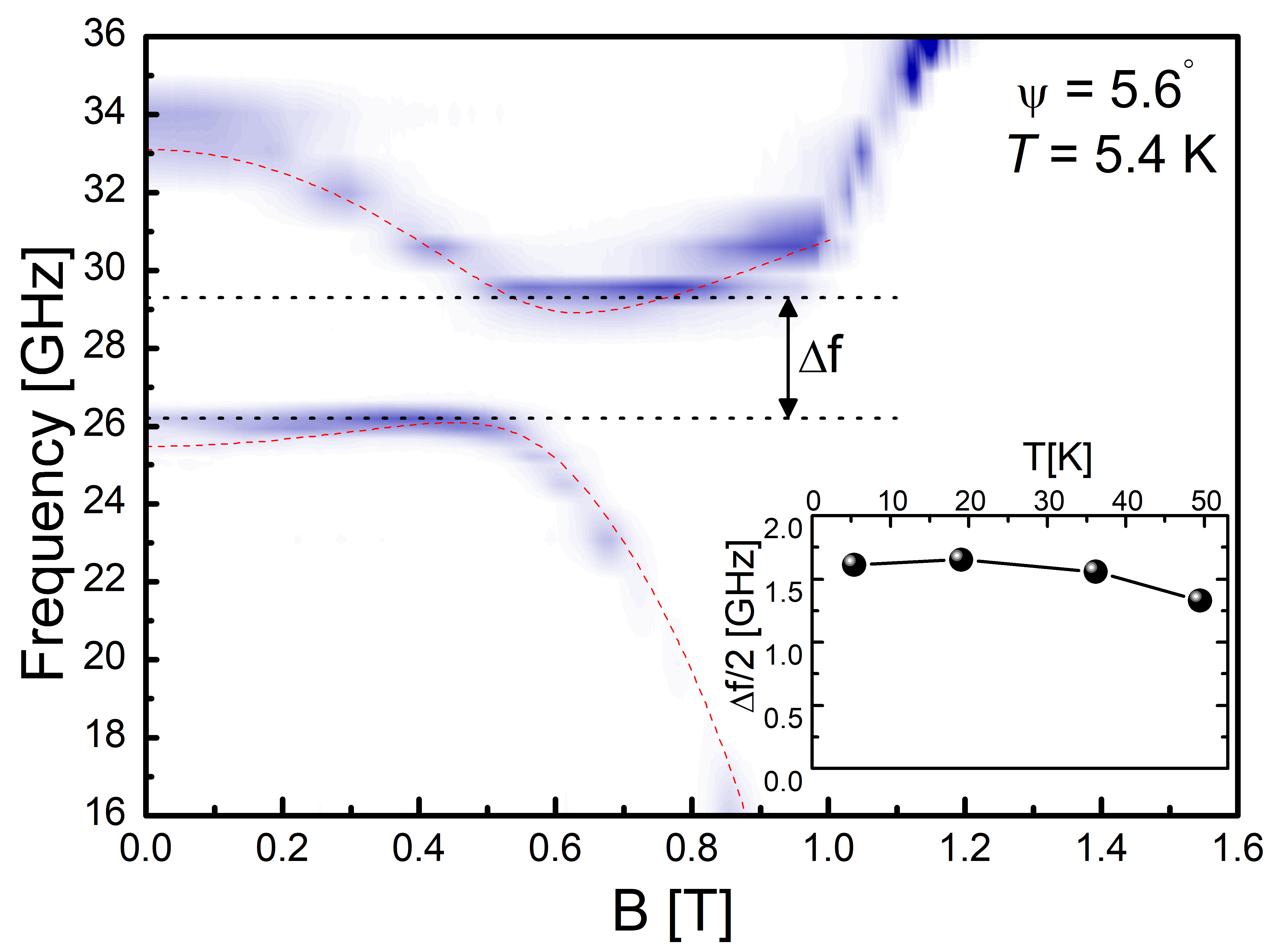}
\end{center}
\caption{(color online) Absorption spectrum for a magnetic field $B$ tilted by an angle $\psi=5.6^{\circ}$ from the a-axis in the (a,b) plane. Color plot of the phase variation of the thermometer photo-resistance as a function of microwave frequency and $B$. The phase difference with respect to its median value (which corresponds to the ``off-resonance'' value) is color-coded from zero (white) up to an upper value of 8\% (dark blue). Theoretical modes (red dashed-lines) obtained with the (fixed) ``global parameters'' of our microscopic model. Inset: coupling strength $\Delta f/2$ for different temperatures (see text for details).}\label{Fig3}
\end{figure}

We note here that a significant modification of the spectrum can be observed with a slight in-plane tilt from $B \parallel$ a-axis towards the b-axis. By introducing a non-zero $B$ component along the (easy) b-axis, the two-fold rotational symmetry of the system is broken and a coupling between magnons can be generated, resulting into the appearance of an anti-crossing behaviour of the two modes (see e.g.  Ref.~\onlinecite{MacNeill2019,Cham2022}). This peculiar regime is reported in the low temperature limit in Fig.~\ref{Fig3}, for an applied magnetic field tilted away from the a-axis by an angle of $\psi=5.6^{\circ}$,  while staying within the (a,b) plane.

At $T=5.4$~K, the minimal frequency spacing between the upper and lower modes is $\Delta f \sim 3.23$ GHz. By defining the coupling strength as $\Delta f/2$, and estimating the resonance frequency linewidths at half maximum $\Gamma_{up}$ and $\Gamma_{low}$ (for the upper and lower mode respectively), one can evaluate the mode cooperativity $C=(\Delta f/2)^{2}/(\Gamma_{up} \times \Gamma_{low})$, which quantify the coupling between magnons. In the present case, at $\psi=5.6^{\circ}$ and $T=5.4$~K, we obtain a cooperativity of $C=8.9$, testifying of a very strong magnon-magnon coupling. The coupling strength $\Delta f/2$ conserves a similar magnitude up to higher temperatures \cite{SM.magnoncoupling}, as can be seen in the inset of Fig.~\ref{Fig3}.

We finally turn to resonance modes for an applied field parallel to the $c$ (hard) axis, i.e. perpendicularly to the sample basal plane (see Fig.~\ref{Fig1}(c)). This dependence is reported as a color map in Fig.~\ref{fig2bis}(c). Both $B=0$ modes now go down with increasing $B$. As the magnetic field is further increased, the saturation of moments along the c-axis is achieved for $B \sim1.8$ T, and a single branch re-emerge for the higher energy mode, with an approximately linear behaviour which can be observed up to the highest studied magnetic field/frequencies. Such a field-dependence was partially reported in  Ref.~\onlinecite{Bae2022}. Finally, we note that throughout our experimental observations, both the upper and lower energy modes may display small internal energy splittings (the order of 0.2--0.3 GHz), clearly identified close to $B=0$ \cite{SM.fielddep}. Other fine features can be observed in the spectra in higher magnetic fields and even above the saturation fields \cite{SM.fielddep}.

To model our experimental observations we develop a microscopic description of CrSBr based on a  general spin Hamiltonian compatible with its orthorhombic symmetry:
\begin{eqnarray}
\hat{\mathcal{H}} & = & \sum_{\langle ij\rangle}
J_{ij}\, \textbf{S}_{i} \cdot \textbf{S}_{j} + \sum_{i} \bigl[ -DS_{i}^{y2}+ E(S_{i}^{z2}-S_{i}^{x2})\bigr]
\nonumber \\
& - &  \sum_i g_\alpha \mu_B B^\alpha S^\alpha \,,
\label{generalH}
\end{eqnarray}
where ${\bf S}_i$ are $S=3/2$ spins of Cr$^{3+}$ ions
and $g_\alpha$ are principal components of the $g$ tensor.
Exchange interactions within the $ab$ layers are predominantly ferromagnetic $J_{ij}<0$ and were determined in the recent inelastic neutron scattering experiment \cite{Scheie2022}. A weak antiferromagnetic exchange  $J_\perp>0$ between layers is responsible for an alternating pattern of layer magnetizations along the $c$ direction in the ordered state below $T_N$. A parameterization of the single-ion anisotropy in Eq.~(\ref{generalH}) is based on the experimental observation that the crystallographic $b$ and $c$ axes correspond, respectively, to easy and hard directions of the magnetization. We note that for an orthorhombic symmetry,
the above parameterization is not unique and one can alternatively use $\hat{\mathcal{H}}_{\rm SI} \simeq [D_cS^{z2}+ E_{ab}(S^{x2}-S^{y2})]$. There is one to one correspondence between various parameterizations of the single-ion term \cite{SM.theory}. Our choice of  $\hat{\mathcal{H}}_{\rm SI}$ in (\ref{generalH}) has the advantage of producing more symmetric expressions for two resonance modes.

Neutron experiments cannot determine  $J_\perp$ and single-ion constants $D$ and $E$ due to insufficient
accuracy at low energies  \cite{Scheie2022}.  Still, the knowledge of these microscopic parameters is essential for establishing magnetic properties of CrSBr in the monolayer limit as well as its potential for information processing. We demonstrate that respective values can be very accurately determined from our resonance measurements. All necessary calculations are performed in the two-sublattice representation of the spin Hamiltonian (\ref{generalH}), which is justified since microwave experiments probe only the long-wavelength magnons.
Intralayer ferromagnetic exchanges do not enter expression for the  ${\bf k} =0$ magnon energies and the exchange term reduces to $\hat{\mathcal{H}}_{\rm ex} = 4J_\perp ({\bf S}_1\cdot {\bf S}_2)$, where the factor 4 reflects the fact that every chromium spin in a buckled $ab$ plane couples to four neighbors  in an adjacent layer situated either above or below the reference layer.
Further details of the calculations are given in the Supplemental Material \cite{SM.theory}  and below we present only the final results.

In zero magnetic field the energies of two resonance modes are given by
\begin{equation}
\Delta_{u,l} = 2S\sqrt{(D\pm E)(4J_\perp + D\mp E)} \,,
\label{zeroBESRmodeu}
\end{equation}
where the upper/lower sign  corresponds  to in-phase/out-of-phase oscillations
of magnetizations in the adjacent layers. In another vdW material CrCl$_3$ with weak antiferromagnetic coupling between ferromagnetic layers, one of the resonance modes has zero frequency due to a negligible anisotropy within the $ab$ plane \cite{MacNeill2019}. Hence, CrSBr is an easy-axis layered material as opposed to the easy-plane CrCl$_3$. This difference has an important effect on the robustness of ferromagnetism down to a single layer of each material.

For magnetic field  along the $b$ axis, CrSBr exhibits an Ising-like transition into a fully polarized state at
$g_b\mu_B B_{\rm cr}^{(b)}= 4J_\perp S$. An intermediate canted phase common in weakly anisotropic antiferromagnets is suppressed by a strong easy-axis anisotropy: $(D-E)> 2J_\perp$ \cite{SM.theory}. In the collinear state at $B < B_{\rm cr}^{(b)}$, the two resonance modes are expressed as
\begin{eqnarray}
\bigl(\Delta_{u,l}/2S\bigr)^2  & = &  4J_\perp D + D^2 - E^2 + \tilde{B}^2
\label{lowBbaxis} \\
&\pm & 2 \sqrt{(2J_\perp E)^2 +  \tilde{B}^2 (4J_\perp D+D^2)}\,,
\nonumber
 \end{eqnarray}
 where $\tilde{B} = g_b\mu_B B/(2S)$. For $B > B _{\rm cr}^{(b)}$, only the upper in-phase mode couples to
a microwave field and its energy is
\begin{equation}
\Delta_u =  2S \sqrt{(D+\tilde{B})^2- E^2}\,.
\label{highBbaxis}
\end{equation}

The above equations give an excellent description of the experimental data as shown in Fig.~\ref{fig2bis}(a). The obtained microscopic parameters satisfy $(D-E) > 2 J_\perp$ confirming a direct `Ising-like' transition at $B_{\rm cr}^{(b)}$ in accordance with the observed resonance frequency jumps. It was argued in literature that at higher temperatures, the $a$-axis anisotropy would be significantly reduced and become \textit{smaller} than the inter-layer exchange, making the intermediate spin-flop phase observable \cite{Cham2022}. At low temperatures, however, our extended magnetic field data \cite{SM.spinflop} show that anisotropy slightly dominates over exchange, explaining the absence of the canted spin-flop phase, consistent with the early conclusion in \cite{Goeser1990}.
The theoretical fit presented in Fig.~\ref{fig2bis}(a) uses the same set of parameters to describe
the excitation spectrum below and above the transition field $B_{\rm cr}^{(b)}$. The corresponding values including
the relevant component of the $g$-tensor are reported in Table~\ref{table1}.

The  resonance  modes can be straightforwardly calculated for two other directions of an applied field.
For $B\parallel a$, the two sublattices tilt continuously towards
 the field by an angle satisfying $\sin\theta = g_a \mu_{B} B/[2S(D-E+4 J_\perp)]$ until
 the critical field $B_{\rm cr}^{(a)}= 2S(D-E+4J_\perp)$ is reached.
In the canted phase for $B < B_{\rm cr}^{(a)}$, the ESR modes are
\begin{eqnarray}
&& (\Delta_u/2S)^2  =  (D+E)(4J_\perp +D-E) \cos^2\!\theta \,,
\label{lowBaaxis}
 \\
&& (\Delta_l/2S)^2 =  (4J_\perp\!+\!D\!+\!E)[4J_\perp \sin^2\! \theta\! +(D\! -\! E) \cos^2\!\theta].
\nonumber
\end{eqnarray}
In the fully polarized state at $B > B_{\rm cr}^{(a)}$, only a single resonance  mode is present with
\begin{equation}
\label{highBaaxis}
\Delta_u = 2S\sqrt{(\tilde{B}+2E)(\tilde{B}-D+E)} \,,
\end{equation}
where $\tilde{B} = g_a \mu_{B} B/2S$. A fit of the full $B$ range ESR data shown in Fig.~\ref{fig2bis}(b) produces
a set of microscopic parameters reported in Table~\ref{table1}.

For  $B\parallel c$, the relevant theoretical expressions are obtained from
Eqs.~(\ref{lowBaaxis}) and (\ref{highBaaxis}) by substitution $E\to -E$ and $g_a\to g_c$.
The corresponding microscopic parameters are given in Table~\ref{table1}.
As can be seen in Fig.~\ref{fig2bis}, a very good description of the \textit{full} 3-axis magnetic field dependence of ESR modes can be obtained with our microscopic model. The parameters obtained from the individual fitting processes for each magnetic field configuration display weak variations (see table~\ref{table1}). This enables us, by using a ``global'' fitting procedure involving our entire multi-axis data set \cite{SM.theory}, to determine a $single$ set of ``global'' microscopic parameters giving an excellent unified description of
the experimental data. The obtained parameters are reported in the last line and column of  table~\ref{table1}.

\begin{table}[h]
\begin{center}
\begin{tabular}{ccccccccc}
\hline\hline $Axis$ & $J_\perp$ (K) & $D$ (K) & $E$ (K)& $B_{\rm cr}$ (T)& $g_\alpha$ & $g_\alpha^{\rm glob}$ \\
\hline
$b$ (easy)             & 0.073 & 0.389  & 0.204 & 0.346 & 1.90  & 1.88  \\
$a$ (intermediate) & 0.069 & 0.399  & 0.215 & 1.015 & 2.03  & 2.05  \\
$c$ (hard)              & 0.068 & 0.387 & 0.181  & 1.90   & 1.97  & 2.04  \\
Global                    & 0.069 & 0.396 & 0.207  & $-$     & $-$    & $-$  \\

\end{tabular}
\end{center}

\caption{CrSBr Microscopic magnetic parameters and saturation fields for the three principal crystallographic directions ($b$ axis (easy), $a$ axis (intermediate), and $c$ axis (hard)).
Inter-layer exchange parameter $J_\perp$, anisotropy parameters $D$ and $E$, $g$-factor and critical fields $B_{\rm cr}$ (see text for definition and \cite{SM.theory} for more details).
Values are extracted for each configuration at $T=5.4$K. The last line and column give the ``global'' parameters giving a common description of the 3 axis magnetic field dependence.} \label{table1}
\end{table}

The presented  description of resonance modes in CrSBr fully agrees and further extends the standard theory
of antiferromagnetic resonance \cite{Nagamiya1955,Gurevich}. The obtained expressions are valid
for arbitrary $D$, $E$, and $J_\perp$. A large value for $D/J_\perp \sim 5.7$ is responsible for a new qualitative effect in CsSBr:
once $B\parallel c$ (hard axis), both resonance modes decrease with increasing field.  Weakly anisotropic
antiferromagnets  with $D\ll J_\perp$ possess  one increasing and one decreasing ESR branch for all field orientations, similar
to what is observed in CrSBr for $B\parallel a,b$.

The microscopic spin model (\ref{generalH}) can be further tested in a tilted magnetic field in the regime of mode repulsion, which is observed already for a small value of the rotation angle  $\psi = 5.6^\circ$ from the $a$ towards the $b$ axis, see Fig.~\ref{Fig3}. Using the `global set' of previously extracted microscopic parameters, we compute for a given $\psi$ the field dependent ESR modes \cite{SM.theory}, and report the results obtained for $\psi = 5.6^\circ$ as red dashed lines in Fig.~\ref{Fig3}. A very good correspondence with the experimental data lays further evidence in the validity of the microscopic model.

The antiferromagnetic inter-plane exchange $J_\perp =0.069$~K is more than 2 orders of magnitude smaller than
the ferromagnetic intra-layer exchanges determined in the inelastic neutron-scattering experiments \cite{Scheie2022}.
Since   $J_\perp$ is also significantly smaller than the single ion anisotropy, we suggest that the transition temperature
in CrSBr is determined by anisotropy together with the in-plane exchange. Further theoretical studies are necessary
to substantiate this proposal. Note, also that $(D-E)=0.188$~K is only modestly larger than $2J_\perp =0.139$~K.
In particular, this means that CrSBr is situated close to a phase boundary separating the Ising-like  versus the spin-flop
 transition for fields along the easy ($b$) axis. This fact can possibly explain why the spin-flop transition was reported at reduced dimensionality in a few-layer sample \cite{Ye2022}, as well as for samples grown under different conditions \cite{Cham2022,Bae2022}.
Such a behavior opens up a possibility to tune magnetic phases by slightly modifying microscopic parameters by external perturbations,
for example, applied strain or pressure.

To conclude, we have reported a detailed magnetic field dependence of the magnon modes in CrSBr by detecting microwave absorptions with a phase-sensitive external resistive detection. Our
data can be very well reproduced by an analytical microscopic model, which leads to the precise determination of the magnetic microscopic parameters of this material. The extracted parameters are consistent
with the weak inter-layer coupling and easy-axis nature of CrSBr, and can be used as a foundation to compute and engineer magnetic phases transitions and devices in this promising material.

We would like to thank C. Bulala, D. Ponton, C. Mollard, J. Spitznagel, I. Breslavetz and K. Paillot for technical assistance.
This work was supported by the Laboratoire d'Excellence LANEF (Grant No. ANR-10-LABX-51-01). CF acknowledges support from the Graphene Flagship. ZS was supported by project LTAUSA19034 from Ministry of Education Youth and Sports (MEYS). MEZ was partly supported by ANR, France, Grant No. ANR-15-CE30-0004.

\pagebreak

\widetext

\begin{center}
\textbf{Microscopic parameters of the van der Waals CrSBr antiferromagnet from microwave absorption experiments: SUPPLEMENTARY INFORMATION}
\end{center}

This supplemental material is organized in five main sections.
Section I gives further details on the experimental approach used and the data acquisition procedure. Section II presents the measurements of $B=0$ magnons.
Section III gives complementary results on the magnetic field dependence of the magnon branches, including results at higher temperature.
Section IV focuses on the magnon-magnon coupling observed by a magnetic-field-induced symmetry breaking.
Finally, section V presents in more details the theoretical approach used to extract microscopic parameters.

\setcounter{equation}{0}
\setcounter{figure}{0}
\setcounter{table}{0}
\setcounter{page}{1}
\makeatletter
\renewcommand{\thefigure}{S\arabic{figure}}

\section{I. Experimental details}

\subsection{A. Samples}

The CrSBr single crystals were synthesized through a chemical vapour transport (CVT) method. Chromium, sulfur and bromine in a stoichiometric ratio of CrSBr were added and sealed in a quartz tube under a high vacuum. The tube was then placed into a two-zone tube furnace. The pre-reaction was done in crucible furnace where one end of ampoule was kept below 300$^{\circ}$ C and the bottom was gradually heated on 700$^{\circ}$ C over a period of 50 hours. Then, the source and growth ends were kept at 800 and 900$^{\circ}$ C, respectively. After 25 hours, the temperature gradient was reversed, and the hot end gradually increased from 850 to 920$^{\circ}$ C over 10 days. High-quality CrSBr single crystals with lengths up to 2 cm were achieved. For the microwave absorption measurements, samples were mechanically exfoliated into thin flakes of typically a few tens of micrometers thick.

\subsection{B. Setup and data acquisition}

Our magneto-absorption experiments were conducted in the GHz regime by monitoring the bolometric response of a resistive sensor placed behind the CrSBr sample, as illustrated in figure 1 in the main text.
Sample and sensor are located in a $^{4}He$ exchange gas environment inside a closed socket, inserted in a variable temperature insert (VTI) fitted into a 16 T superconducting magnet. The VTI Temperature was PID regulated and the effective temperature $T$ of the sample under microwave radiation was measured via the calibrated sensor resistance $R$.

To enhance the measurement sensitivity and minimize thermal drifts, a low frequency modulation of the microwave was employed, and the thermometer response was measured with a standard low frequency AC lock-in technique, monitoring both its amplitude $\Delta R$ and phase $\phi$. The thermometer AC voltage response is generally anti-phased ($\Delta R<0$) with respect to the microwave modulation reference, due to the insulator-like behaviour of the sensor (R decreases as the temperature increases). In practice, there is a finite thermalisation time for the thermometer, which makes the actual phase of the signal different from $\pi$. As a microwave absorption takes place in the sample, an increase in the modulus $|\Delta R|$ of the photo response signal reveals a lower effective temperature of the thermometer (since the first derivative of $R(T)$ increases at lower $T$), consistent with the fact that a fraction of the incoming power has been absorbed by the sample. The phase of the signal also undergoes a change during the absorption process, which we interpret as a change in thermalisation times, again due to the sample absorption.

Measurement were generally (but not only, see section II) performed by sweeping the magnetic field at a fixed microwave frequency and temperature.  To perform a study at a fixed temperature (which if not constant, could cause a shift of the lines in case of a strong temperature dependance of the magnon dispersion), the microwave generator output power has been slightly adjusted from one frequency to another to target a similar thermometer average temperature. This enables to compensate both for the generator and microwave circuit frequency-dependent power output and transmission (the influence of the microwave power on the signal is additionally discussed in section I.C. The size of the frequency steps used to build the frequency/magnetic field spectra shown in the main text was chosen as a function of the desired resolution: for the b-axis configuration, the frequency steps vary from 1 to 6 GHz depending on the region of interest, for the a-axis configuration, from 0.25 to 2.5 GHz, and for the c-axis configuration, from 2 to 5 GHz. The color plots in the main text figure are made from a linear interpolations of the phase of the thermometer photo-resistance between the fixed frequency scans.

To further confirm the resonant origin of the observed changes in the signal, data were systematically taken in a symmetric magnetic field range, up to 10 T. The magnetic field sweep rate was chosen to be small enough (typically 0.2 T per minute) to have no influence of the resonances lineshape, and the latter was also found to be insensitive to the sweep direction.

\subsection{C. Power dependance}\label{power}

\begin{figure}[!h]
\begin{center}
\includegraphics[width=12cm]{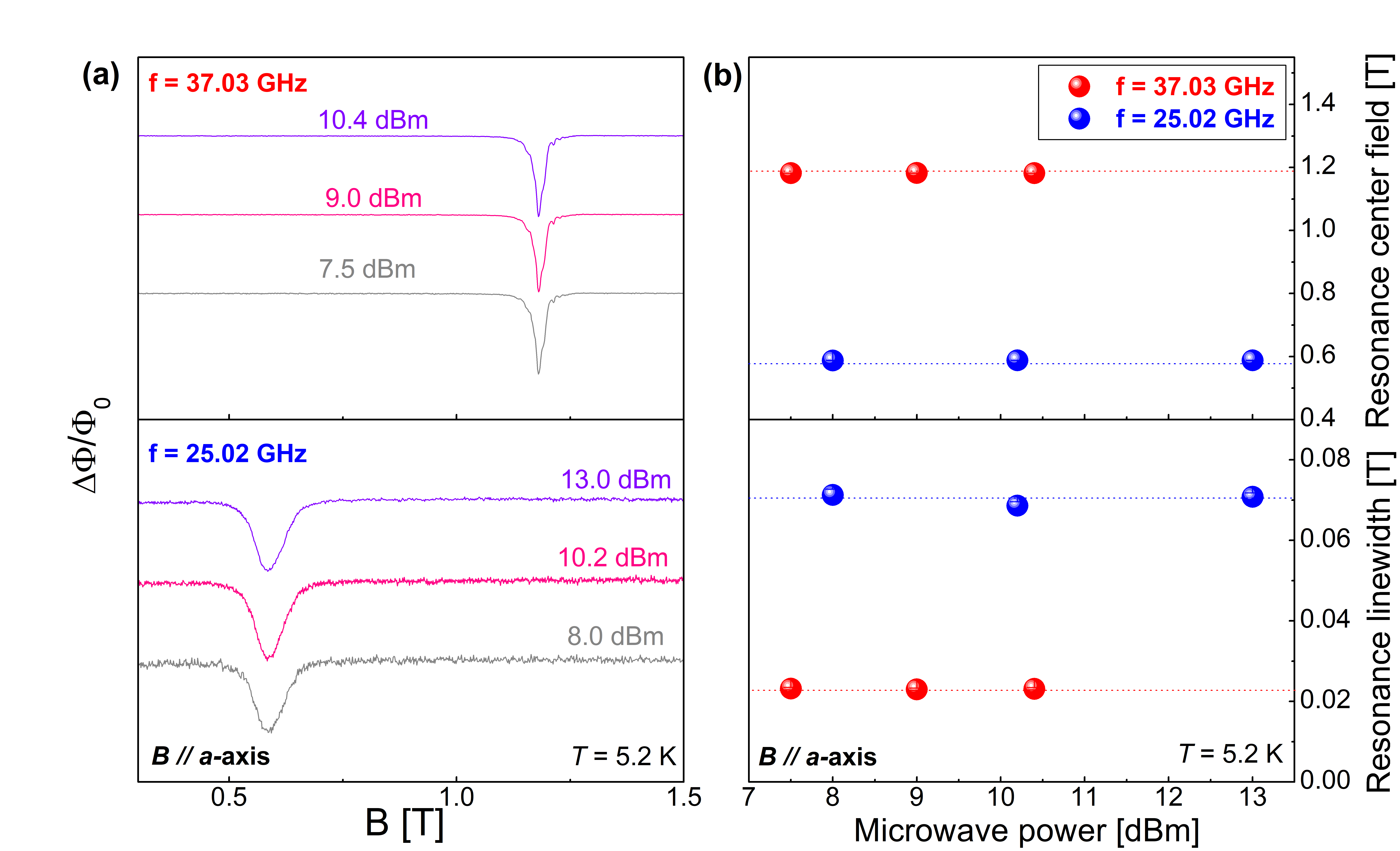}
\end{center}
\caption{(color online) Microwave power dependence of the magnon excitations. Left-panel (a): relative change (arbitrary Y-scale) of the phase of the thermometer photo-resistance $\phi$ with respect to its ``off-resonant'' value $\phi_{0}$ as a function of the magnetic field $B$ (in this example, applied along the a-axis), for different microwave output powers (from 7.5 to 13 dBm). The top (bottom) panel displays resonances of the upper (lower) energy mode, and microwave powers are indicated above each traces. Right-panel (b): Resulting center fields (top panel) and linewidths (Full width at half maximum, bottom panel) of the resonant signals, with the horizontal dotted lines corresponding to the average values. Red (blue) dots correspond to the upper (lower) energy mode.
}\label{Figpower}
\end{figure}

The effect of the microwave excitation power on the resonant signals was studied for a fixed bath temperature, and is reported in Fig.~\ref{Figpower}.
Our results show that in the range of power used (from a minimum of 7.5 dBm to a maximum of 13 dBm at the generator output), there were no power-related changes of the lineshape, linewidth and position (center field) of the resonance. This implies, in particular, that microwave heating effects on the sample at high power were not significant (and remained anyway under control thanks to the determination of the effective temperature under radiation via our resistive thermometer).

\section{II. $B=0$ magnon energies}\label{zeroBmagnon}

\begin{figure}[]
\begin{center}
\includegraphics[width=9cm]{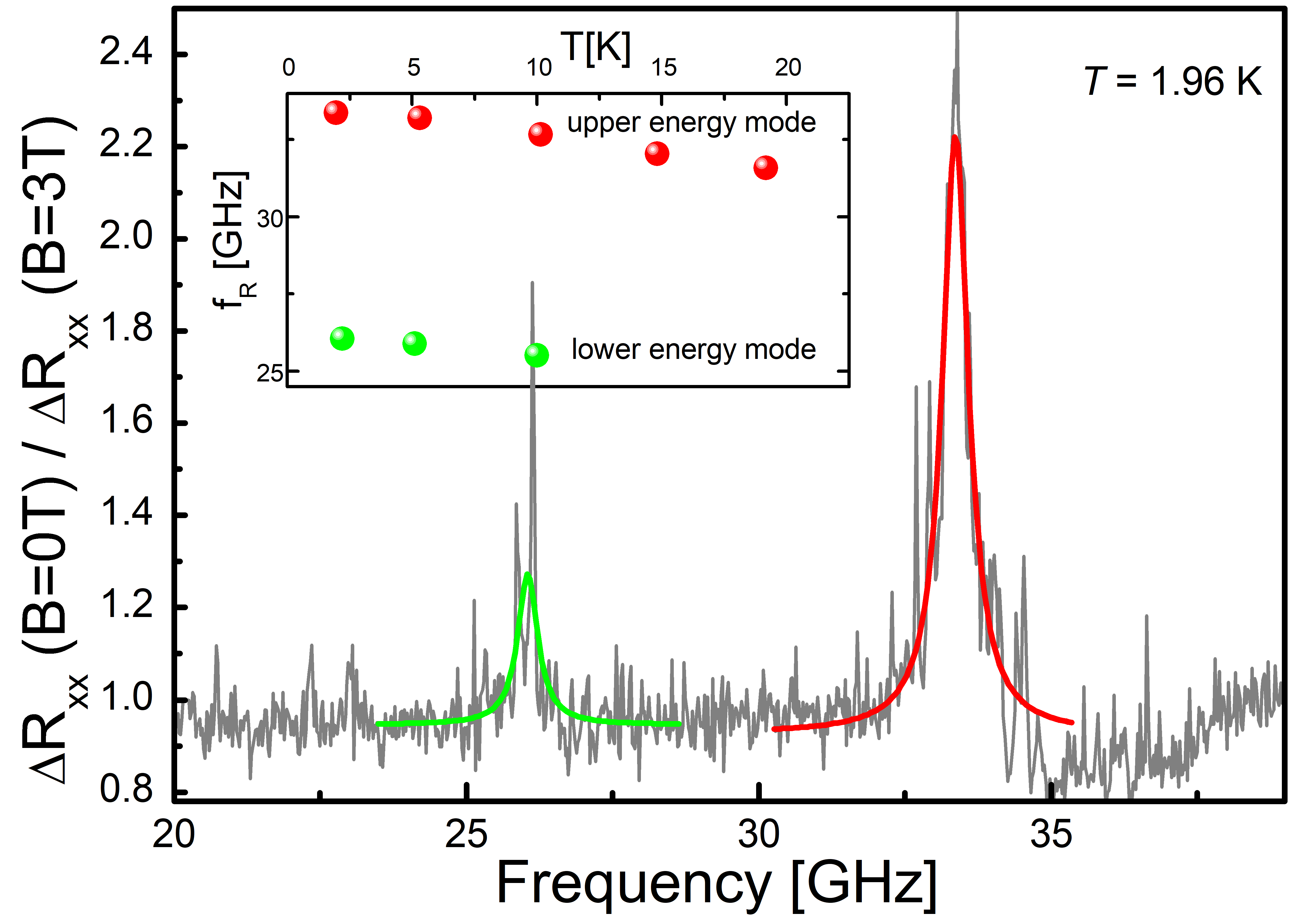}
\end{center}
\caption{(color online) Determination of the zero-field magnons energies with frequency sweeps. Normalized relative change of the thermometer photo-resistance $\Delta R$ (solid grey line) as a function of the microwave frequency. The normalization is made by dividing the $B=0$ frequency sweep by a frequency sweep taken at a different finite magnetic field (here $B=3$ T). Solid green and red lines are Lorentzian fits of the resonances, from which the
resonant frequency $f_{R}$ is extracted. Inset: temperature dependence of the zero-field resonant frequencies for each mode. Red (green) dots correspond to upper (lower) frequency mode.
}\label{freqsweep}
\end{figure}

The $B=0$ magnon energies were extracted with \textit{frequency} sweeps, an example of which is given in Fig.~\ref{freqsweep}. Due to the non-negligible power fluctuations of the microwave source as a function of frequency,
this sweeping mode generally displays significantly more noise and signal fluctuations compared to the magnetic field sweeps. Nevertheless, by normalizing the $B=0$ sweep with a frequency sweep taken at different fixed magnetic field (where resonant features have moved sufficiently in frequency), the resonant frequencies emerge, as can be seen in Fig.~\ref{freqsweep}. The extracted values at our lowest studied temperature ($T=1.96$ K) are in this case: $26.05 \pm 0.01 $ GHz  and  $33.37 \pm 0.01$ GHz for the lower and upper modes, respectively.

As observed for the magnetic-field sweeps (see main text), the resonance linewidths are rather small: the full width at half maximum (FWHM) are 0.42 GHz and 0.55 GHz for the lower and upper mode respectively, confirming both the sensitivity of our measurements and the good sample quality. Note that these FWHM are actually ``boosted'' by internal splittings, described in the following section III.A. This experiment was repeated at difference temperatures, showing a decrease of the resonant frequencies of both modes as the temperature increases (see the inset of Fig.~\ref{freqsweep}). This decrease is associated with the weakening of anti-ferromagnetism with temperature and consistent with other experimental observations \cite{Bae2022}.

We finally note that the zero-field magnons energy can also be extracted with magnetic field sweeps (as shown in the main text), but because of the local flatness of the $B$-dispersion close to $B=0$~T, resonant signals can be quite broad in $B$-sweeps and the resonant frequency extraction becomes a bit less precise.

\section{III. Field-dependent magnon dispersion}

\subsection{A. Extended data} \label{extended}

\begin{figure}[h]
\begin{center}
\includegraphics[width=14cm]{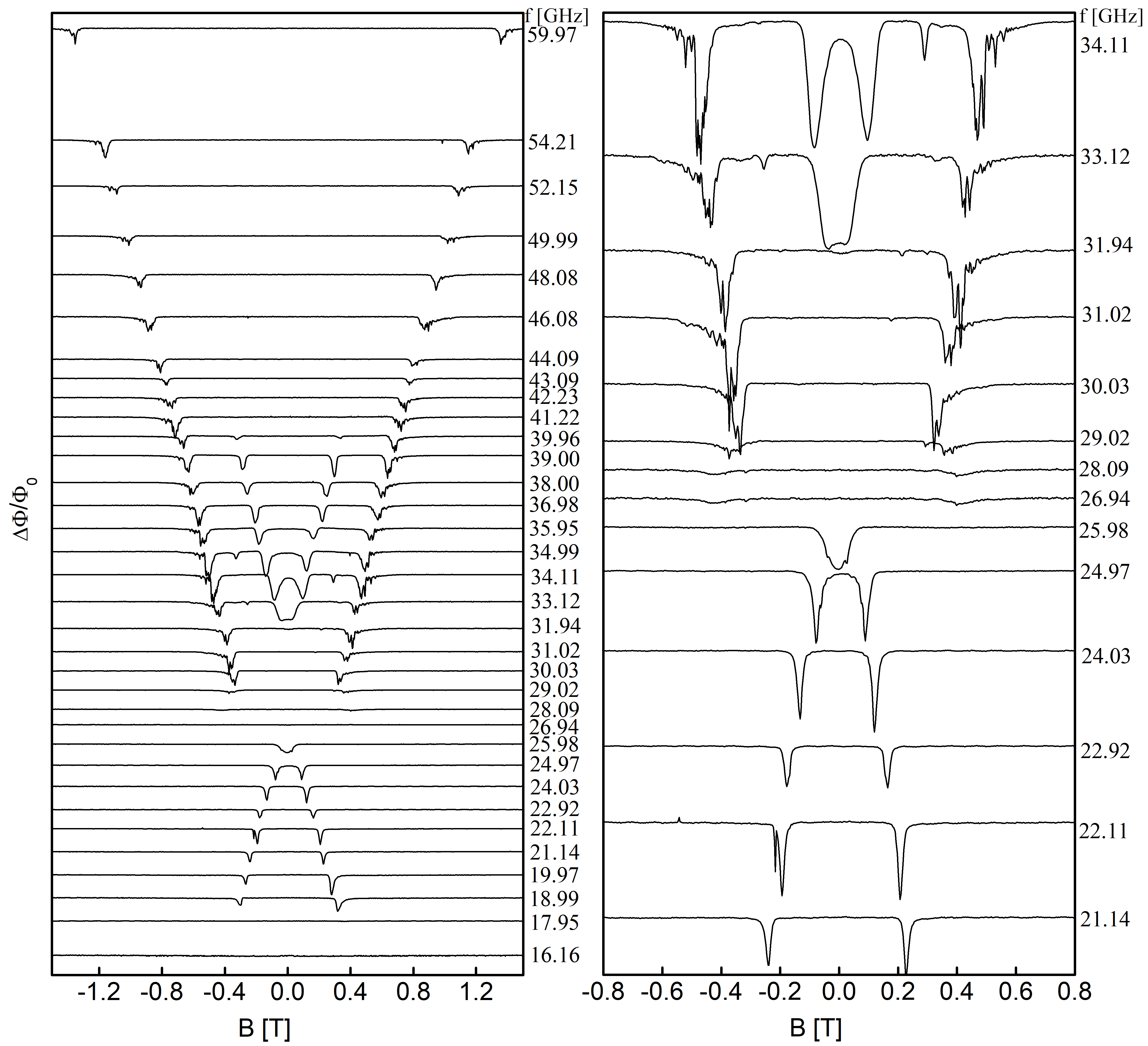}
\end{center}
\caption{(color online) Waterfall plots of relative variations of the thermometer phase photo-response for a a magnetic field $B$ applied along the b-axis (arbitrary Y-scale). Left panel: Selected data on the entire studied range. Right panel: zoom-in data closer to the $B=0$ magnon energies.
}\label{waterfallb}
\end{figure}

\begin{figure}[h]
\begin{center}
\includegraphics[width=14cm]{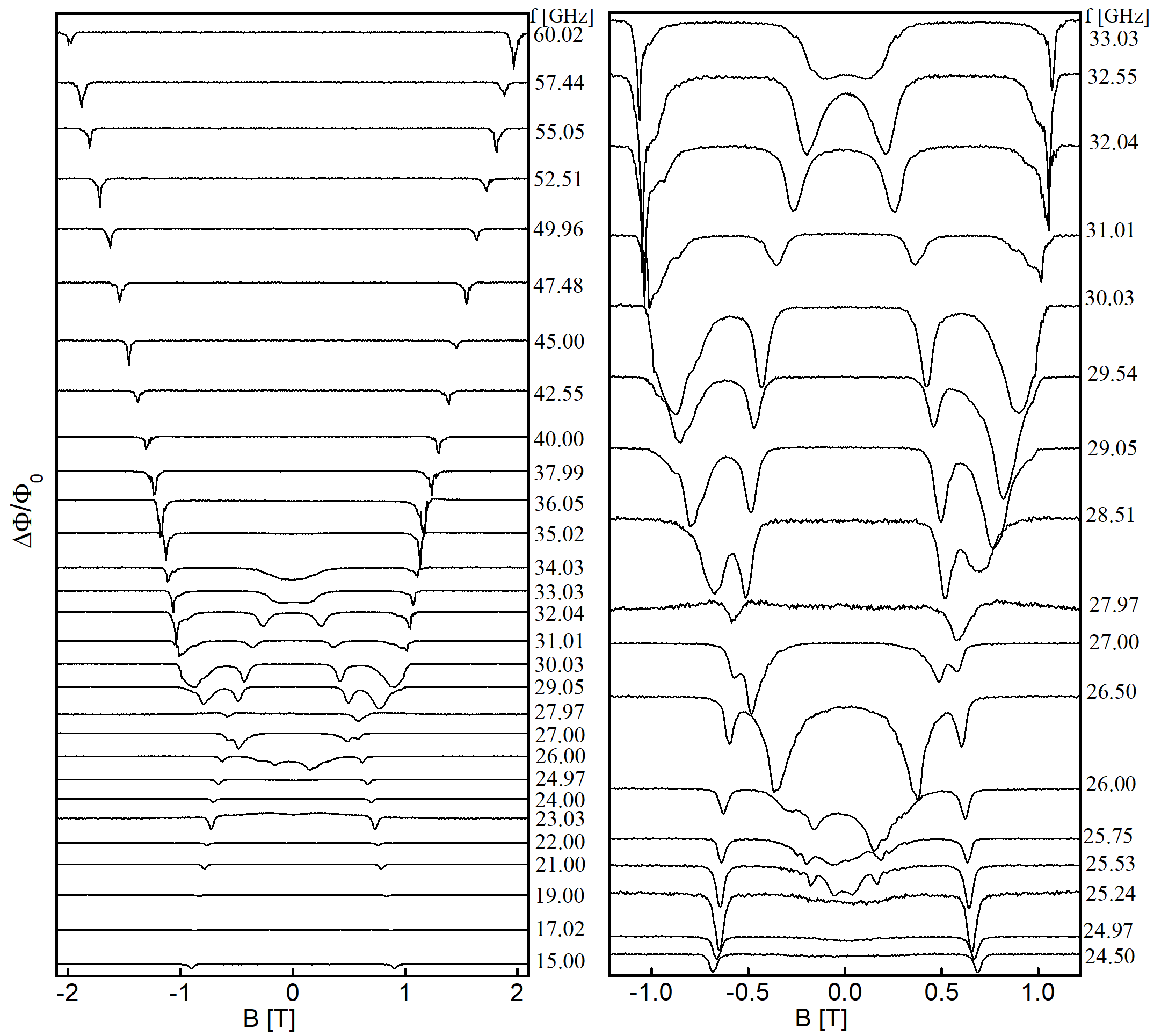}
\end{center}
\caption{(color online) Waterfall plots of relative variations of the thermometer phase photo-response for a a magnetic field $B$ applied along the a-axis. Left panel: Selected data on the entire studied range (arbitrary Y-scale). Right panel: zoom-in data closer to the $B=0$ magnon energies.
}\label{waterfalla}
\end{figure}

\begin{figure}[h]
\begin{center}
\includegraphics[width=14cm]{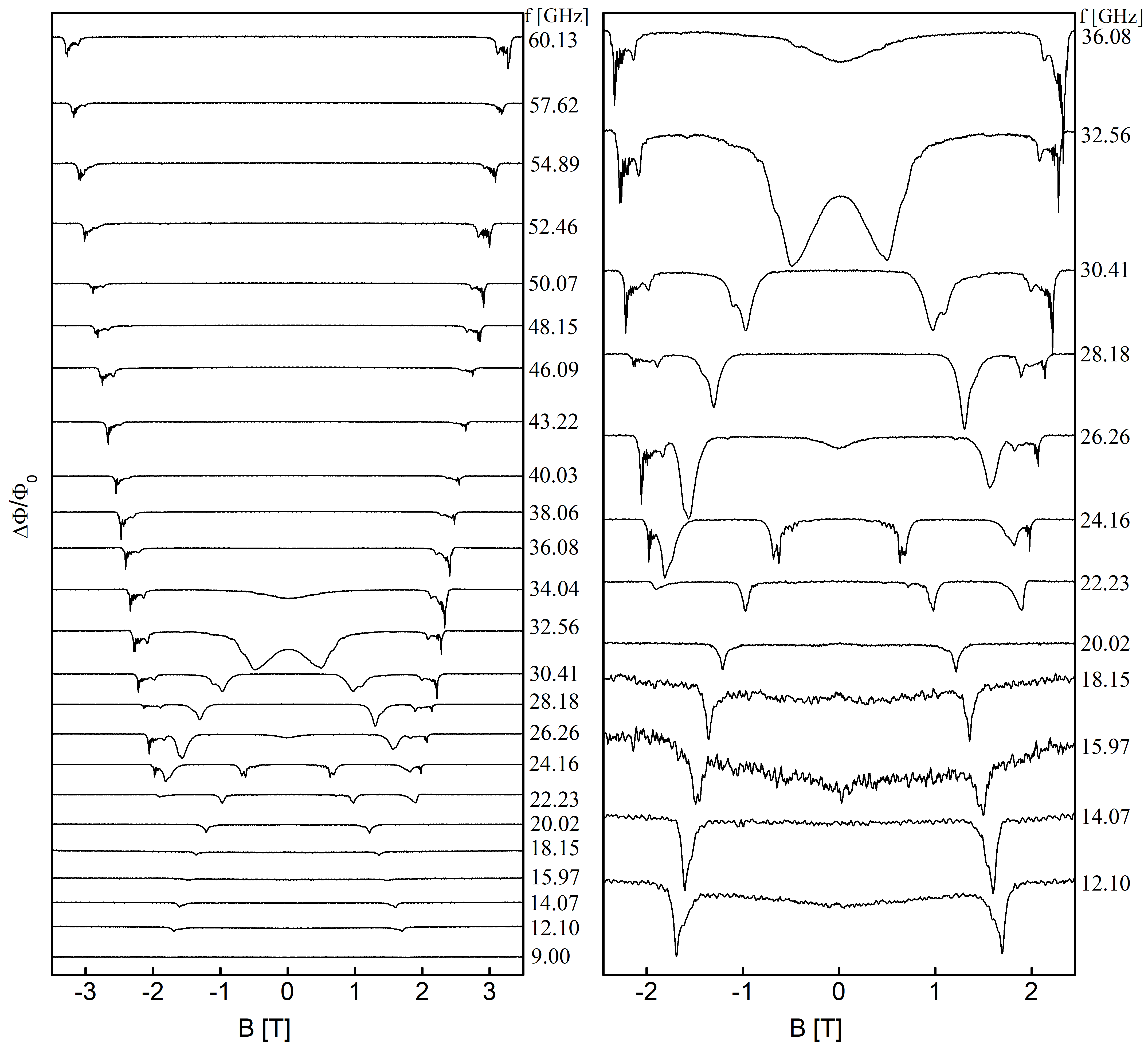}
\end{center}
\caption{(color online) Waterfall plots of relative variations of the thermometer phase photo-response for a a magnetic field $B$ applied along the c-axis. Left panel: Selected data on the entire studied range (arbitrary Y-scale). Right panel: zoom-in data closer to the $B=0$ magnon energies.
}\label{waterfallc}
\end{figure}

Figure ~\ref{waterfallb}, figure ~S\ref{waterfalla} and figure ~\ref{waterfallc}, show raw data as waterfall plots for the b-axis, a-axis, and c-axis.
These data can reveal finer features. First, we can observe small internal splittings of the main resonances close to $B=0$, regardless of the applied magnetic field configurations. For the lower energy mode, these are
for example very clear on the scans at $f=25.98$~GHz along the b-axis, $f=25.53$~GHz along the a-axis, or $f=24.16$~GHz along the c-axis. Similar splittings, somewhat much less pronounced, can be seen for the upper
energy mode (for example at $f=33.03$~GHz along the a-axis). These splittings can actually also be seen in the frequency sweeps of Fig.~\ref{freqsweep}, in particular for the lower energy mode,
and could be related to a hybridization between magnon and phonon modes \cite{Bae2022}.

Some of them seem to be traceable under higher magnetic fields even though the magnon energy is modified. For example, for $B$ along the c-axis, clear splittings are seen between 1 and 3 T both for the lower
and upper mode. See for example the scans at 15.97 and 14.07~GHz for the former and the one at 30.41~GHz for the latter, where a double dip-like response can be seen even above the saturation field (and up to 60~GHz).

For $B$ along the b-axis, we note that a peculiar behaviour is observed around $ B_{\rm cr}^{(b)}$  (with downward dispersing lines in the 26.94~GHz-29.02~GHz region) suggesting different magnetic domains could be formed in the region of transition to the ferromagnetic state. The existence of domains near the critical field is actually clearer in the a-axis data: in the 29-31~GHz region, 2 responses coexist. A lower $B$ response corresponds to the canted antiferromagnetic system and follows the behaviour expected for  $ B < B_{\rm cr}^{(a)}$, and a response at slightly higher $B$ which corresponds to the linear in $B$ dispersion expected for $ B > B_{\rm cr}^{(a)}$, where the magnetization is fully saturated. The different dispersions of those two field regimes give rise to a different linewidth, and we observe a broad low $B$ shoulder and sharp minimum on the higher $B$ side of the resonance.

The resolution of these fine features further highlight the sensitivity of our technique. The discussion of the detailed physics behind them is nevertheless outside the scope of the present paper.

\subsection{B. Effect of temperature}

\begin{figure}[!h]
\begin{center}
\includegraphics[width=9cm]{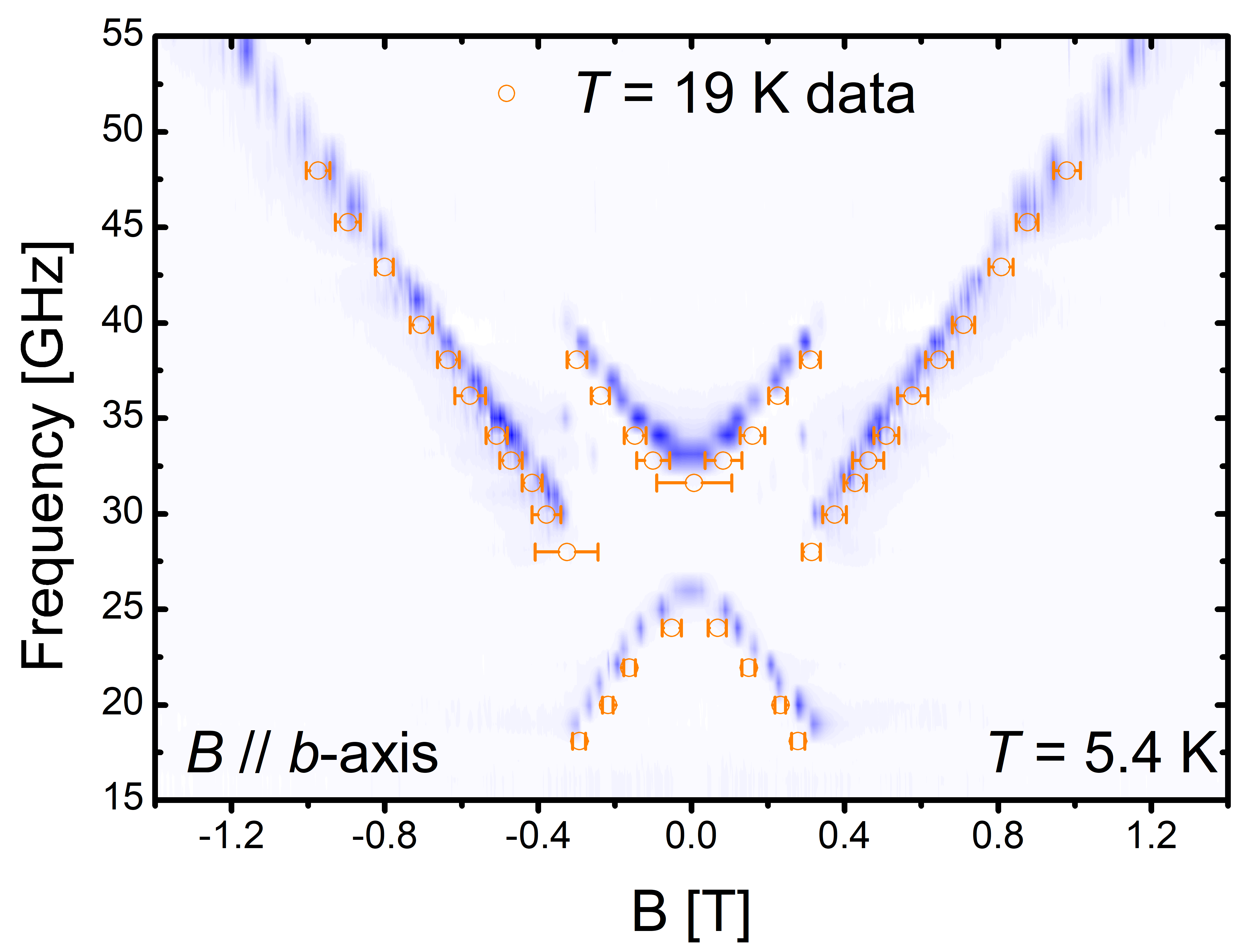}
\end{center}
\caption{(color online) Absorption spectrum for a magnetic field $B$ along the b-axis. Color plot of the phase variation of the thermometer photo-resistance (same data as main text) as a function of microwave frequency and magnetic field $B$ along the b-axis, for $T=5.4$~K. Data at a higher temperature of $T=19$~K are plotted as orange open circles.
}\label{Tdepbaxis}
\end{figure}

The magnon dispersion was measured for $H\parallel$ b-axis at a higher temperature of about $T=19$~K, and is reported in Fig.~\ref{Tdepbaxis}. The global downshift of the resonant frequencies is initiated by the reduction
of the zero-magnetic field resonant frequencies with temperature (see section II). Additionally, the critical field $ B _{c}^{b}$ is reduced at higher temperature which is attributed to the thermal weakening of anti ferromagnetic interactions.

\section{IV. Generation of a strong magnon-magnon coupling}

As described in the main text, and observed at higher temperature in Ref.~\onlinecite{Cham2022}, a strong magnon-magnon coupling can be generated by introducing a non-zero $B$ component along the (easy) b-axis, starting from the $B\parallel$ a-axis configuration. This results into the appearance of an anti-crossing behaviour of the two modes, similar to the one previously observed in CrCl$_{3}$ \cite{MacNeill2019} in another symmetry breaking configuration. For an angle of $\psi=5.6^{\circ}$ between $B$ and the a-axis, we have observed this effect in the low temperature limit where a strong coupling strength $(\Delta f/2)$=1.615 GHz ($\Delta f$ is the minimal frequency spacing between the upper and lower modes) was observed. By estimating the resonance frequency linewidths at half maximum $\Gamma_{up}$ and $\Gamma_{low}$ (for the upper and lower mode respectively), one can evaluate the mode cooperativity $C=(\Delta f/2)^{2}/(\Gamma_{up} \times \Gamma_{low})$, which quantifies the coupling between magnons. In the present case, at $\psi=5.6^{\circ}$ and $T=5.4$~K, we obtain a cooperativity of $C=8.9$. This mode cooperativity, higher than the previously reported ones in CrSBr or CrCl$_{3}$, can be tuned by the $\psi$ angle value and, in principle, even further increased by improving the sample quality. This makes CrSBr an interesting platform to study magnon-magnon interactions.

\begin{figure*}[]
\begin{center}
\includegraphics[width=18cm]{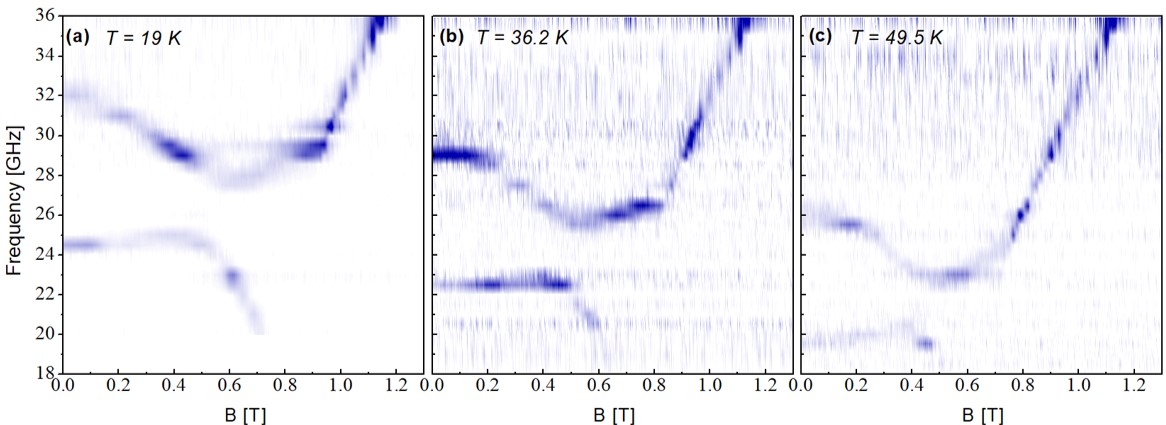}
\end{center}\caption{(color online)  Absorption spectra for a magnetic field $B$ tilted by an angle $\phi=5.6 ^{\circ}$ from the a-axis in the (a,b) plane, for different temperatures. Color plot of the phase variation of the thermometer photo-resistance as a function of microwave frequency and $B$. The phase difference with respect to its median value (which corresponds to the ``off-resonance'' value) is color-coded from zero (white) up to an upper value of 8\% (dark blue).}\label{anticrossingT}
\end{figure*}

We have investigated the becoming of the mode anti-crossing in higher temperatures, with some examples being reported in Fig.~\ref{anticrossingT}. As can be seen, the coupling strength is not much affected by temperature up to 50 K.

\section{V. Theoretical approach and microscopic parameters }

\subsection{A. Spin model}

The minimal spin model for CrSBr includes isotropic exchange interactions between neighboring $S=3/2$ spins
of chromium ions and the single ion anisotropy term compatible with the orthorhombic crystal symmetry:
\begin{equation}
\hat{\mathcal{H}} = \sum_{\langle ij\rangle} J_{ij}\, \textbf{S}_{i} \cdot \textbf{S}_{j} +
\sum_{i} \bigl\{ -D (S_i^{y_0})^2+ E\bigl[ (S_i^{z_0})^2- (S_i^{x_0})^2\bigr]
- g_\alpha \mu_B B^\alpha S^\alpha_i\bigr\}.
\label{H0}
\end{equation}
We have also included a Zeeman term with an anisotropic $g$-tensor,  $g_\alpha$ being its principal components.
Negative exchange interactions within $ab$ layers  $J_{ij}<0$ are responsible for the
formation of uniform ferromagnetic polarization in each layer. The positive interlayer coupling
 $J_\perp>0$ produces antiferromagnetic stacking of ferromagnetic layers  along the $c$ axis.

We note that for an orthorhombic symmetry, the above parameterization is not unique and one can alternatively use $\hat{\mathcal{H}}_{\rm SI} \simeq [D_cS^{z2}+ E_{ab}(S^{x2}-S^{y2})]$. There is one to one correspondence between various parameterizations of the single-ion term (see e.g. section V.C. and table \ref{table2} below). Our choice of  $\hat{\mathcal{H}}_{\rm SI}$ in (\ref{H0}) has the advantage of producing more symmetric expressions for two resonance modes.

The resonant ESR spectra in an antiferomagnetic state can be obtained in
the two-sublattice representation of the total spin Hamiltonian (\ref{H0}). For CrSBr we thus have
\begin{equation}
\hat{\mathcal{H}} =  4J_\perp\, \textbf{S}_1 \cdot \textbf{S}_2 +
\sum_{i=1,2}  \bigl\{ -D (S_i^{y_0})^2+ E\bigl[ (S_i^{z_0})^2- (S_i^{x_0})^2\bigr] -
 g_\alpha \mu_B B^\alpha S^\alpha_i\bigr\}.
\label{H}
\end{equation}

For $B\parallel b$ (easy) axis, the classical energy ($T=0$) of a collinear antiferromagnetic state is $E_{\uparrow\downarrow} = -4J_\perp S^2 -2DS^2$.
It remains field independent due to vanishing magnetization. The energy of a fully polarized state, which appears in strong magnetic fields,
is $E_{\uparrow\uparrow} = 4J_\perp S^2 -2DS^2 - 2g_b\mu_BBS$. Another magnetic state that may intervene between the two above states is
a canted antiferromagnetic state. In this state the opposite magnetic sublattices are oriented along the intermediate $a$ ($x$) axis
with uniform tilting angle $\theta$  along the $b$ direction. The corresponding classical energy is
\begin{equation}
E_{\rm CAF} = - 4J_\perp S^2\cos 2\theta - 2ES^2\cos^2\theta - 2DS^2\sin^2\theta - 2g_b\mu_B BS\sin\theta \,.
\end{equation}
minimization with respect to $\theta$ yields
\begin{equation}
\sin\theta = \frac{g_b\mu_B B}{2S(4J_\perp+E-D)}\,.
\end{equation}
Hence,
\begin{equation}
E_{\rm CAF} = - 4J_\perp S^2 - 2ES^2 - \frac{(g_b\mu_B B)^2}{2(4J_\perp + E-D)} \,.
\end{equation}

Comparing  $E_{\rm AF}$, $E_{\rm CAF}$, and $E_{\rm FM}$ we obtain the following sequence of phase transitions.
For small anisotropy $(D-E)<2J_\perp$, the sequence of phases during a magnetizaion process is
$\uparrow\downarrow \;\to\; {\rm CAF}\; \to\; \uparrow\uparrow$ with two transition fields in between
\begin{equation}
g_b \mu_B B_{\rm sf} =  2S\sqrt{(D-E)(4J_\perp -D+E)} \,, \qquad
g_b \mu_B B_{\rm sat} =  2S(4J_\perp -D+E)\,.
\end{equation}
The two transition fields merge at $(D-E)=2J_\perp$ and for larger anisotorpy $(D-E)>2J_\perp$ there is
a direct Ising-like transition from antiferromagnetic to fulle saturated state,
$\uparrow\downarrow \;\to\; \uparrow\uparrow$ at
\begin{equation}
g_b \mu_B B_{\rm cr}^{(b)} = 4J_\perp S \,.
\end{equation}

For orthogonal field orientations spins start to tilt from the $b$ axis already by small magnetic
fields. For example, for $B\parallel a$ the tilting angle is
\begin{equation}
\sin\theta = \frac{g_a\mu_B B}{2S(4J_\perp + D - E)}\,.
\end{equation}
Sublattices become fully aligned with the field ($\theta =\pi/2$) at the critical field
\begin{equation}
g_a \mu_B B_{\rm cr}^{(a)} =  2S(4J_\perp + D - E) \,.
\end{equation}

\subsection{B. Spin-wave calculations}

In the following, we represent spins via boson operators using the standard Holstein-Primakoff transformation
\cite{Holstein1940}.
All subsequent calculations are performed in the harmonic approximation, hence it suffices to write
\begin{equation}
S^z_i = S - a^\dagger_ia_i \,, \qquad S_i^x \approx \sqrt{\frac{S}{2}}\,(a^\dagger_i + a_i)\,, \qquad
S_i^y \approx \sqrt{\frac{S}{2}}\,\frac{1}{i}\,(a^\dagger_i - a_i)\,.
\label{HP}
\end{equation}

Another important point is that the Holstein-Primakoff transformation must be applied
in the local frame $(x_i,y_i,z_i)$, $i=1,2$,  assossiated  with each antiferromagnetic sublattice,
whereas the anisotropic terms in (\ref{H0}) and (\ref{H}) are written in the global
crystallographic frame $(x_0,y_0,z_0)$.

Let us consider in detail how the above steps are performed in zero field.  In this case
the antiferromagnetic sublattices  are oriented up and down along the easy axis ($y_0$). The spin
Hamiltonian (\ref{H}) is written in the local frame as
\begin{equation}
\hat{\mathcal{H}} =  4J_\perp\,\bigl( S_1^y S_2^y  -  S_1^z S_2^z -  S_1^x S_2^x\bigr)
- D \bigl[(S_1^z)^2 + (S_2^z)^2\bigr] + E\bigl[
(S_1^y)^2 + (S_2^y)^2 - (S_1^x)^2 + (S_2^x)^2\bigr].
\label{Hloc}
\end{equation}

Substitutting the Holstein-Primakoff transormation and keeping only the quadratic terms we obtain
\begin{equation}
\hat{\cal H}_2  =  4J_\perp S \bigl(a^\dagger_1 a_1 + a^\dagger_2 a_2  - a_1 a_2
- a^\dagger_2 a^\dagger_1 \bigr) + 2DS \bigl(a^\dagger_1 a_1 + a^\dagger_2 a_2\bigr)
- ES \bigl(a_1^2 + a_2^2 + a^{\dagger 2}_1 + a^{\dagger 2}_2\bigr).
\label{H2}
\end{equation}

Diagonalization of the quadratic bosonic Hamiltonians (\ref{H2}) that contain both normal and anomalous terms
is performed witht the help of the generalized Bogolyubov transformation
\begin{equation}
b_n = u_1 a_1 + u_2 a_2 + v_1 a_1^\dagger + v_2 a_2^\dagger \,.
\end{equation}
The coefficients $u,v$ are chose in such a way that new operators $b_n$ staisfy the bosonic ommutatioon
relations  $[b_n, b_n^\dagger] = 1$ and diagonalize the Hamiltonian
\begin{equation}
\hat{\cal H}_2  = \sum_n \omega_n b_n^\dagger b_n \quad \textrm{or}\quad
\omega_n b_n = [b_n,\hat{\cal H}_2] \,.
\label{Hdiag}
\end{equation}

The diagonalization condition leads to the matrix equation on the $u,v$ coefficients:
\begin{equation}
\omega \left( \begin{array}{c} u_1 \\ u_2 \\ v_1 \\ v_2 \end{array}\right) =
S\left( \begin{array}{cccc}
2D + 4J_\perp &  0  &  2E  &  4J_\perp \\
  0 & 2D + 4J_\perp & 4J_\perp  &  2E  \\
-2E &  -4J_\perp & -2D-4J_\perp & 0 \\
 -4J_\perp & -2E & 0 & -2D - 4J_\perp
\end{array}
\right)
\left( \begin{array}{c} u_1 \\ u_2 \\ v_1 \\ v_2 \end{array}\right) .
\label{Dynam}
\end{equation}
The eigenvalue problem (\ref{Dynam}) leads to a biquadratic equation with the positive roots
given by
\begin{equation}
\omega_{1} = 2S\sqrt{(D + E)(4J_\perp + D - E)} \,, \qquad
\omega_{2} = 2S\sqrt{(D - E)(4J_\perp + D + E)} \,.
\end{equation}
which coinsides with the expression given in the text.

In an applied magnetic field spins gradually tilt by angle $\theta$ towards the field direction. For example,
for $B\parallel x_0$ ($a$) axis the transformation between the global and the local frames goes as
\begin{eqnarray}
&& S_1^{x_0} = S_1^z \sin\theta + S_1^x \cos\theta \,, \quad S_1^{y_0} = S_1^z \cos\theta - S_1^x \sin\theta \,,
\quad \ \ S_1^{z_0} = S_1^y \,,
\nonumber \\
&& S_2^{x_0} = S_2^z \sin\theta - S_2^x \cos\theta \,, \quad S_1^{y_0} = -S_1^z \cos\theta - S_1^x \sin\theta \,,
\quad S_2^{z_0} = S_2^y \,.
\label{CAF}
\end{eqnarray}
Substitutting (\ref{CAF}) into (\ref{H}) and treating spins as classical vectors ($S_i^z\to S$)
we obtain the energy of the canted state as
\begin{equation}
E = - 4J_\perp S^2\cos 2\theta - 2DS^2\cos^2\theta - 2ES^2\sin^2\theta - 2g_a\mu_B BS\sin\theta \,.
\end{equation}
minimization with respect to $\theta$ yields
\begin{equation}
\sin\theta = \frac{g_a\mu_B B}{2S(D-E+4J_\perp)}\,.
\end{equation}
Sublattices become fully aligned with the field ($\theta =\pi/2$) at the critical field
\begin{equation}
g_a \mu_B B_{\rm cr}^{(a)} =  2(D-E+4J_\perp)S \,.
\end{equation}
Calculation of the resonance spectra is performed by keeping all spin components in the transformed Hamiltonian
and, then, performing steps similar to what was done for the $B=0$ case.

Finally, we present here frequencies of resonance modes in a tilted field, which have been omitted from
the main text due to their length. Assuming a magnetic field rotated by a small angle $\psi$ from the $a$ towards the $b$ axis
we obtain from the spin wave calculations
\begin{equation}
\Omega_{1,2} = \frac{\omega_1+\omega_2}{2} \pm \sqrt{ \frac{(\omega_1-\omega_2)^2}{4} + \Lambda^2} \,,\quad
\Lambda = C_1 (u_1 u_2 + v_1 v_2) + C_2 (u_1 v_2 + u_2 v_1) \,.
\label{Lambda}
\end{equation}
Here $\omega_{1,2}$ coinside with $\Delta_{u,l}$ given by Eq.~(5) in the main text, whereas
\begin{equation}
 C_1 = \mu_B B \cos\theta (g_b \psi - g_a\varphi) + 3C_2 \,, \quad
 C_2 = (D-E)S\varphi\,\sin2\theta \,,\quad
 \varphi = \psi\,\frac{g_b \mu_B B \sin\theta}{g_a \mu_B B \sin\theta + 2(D-E)S\cos 2\theta} \,.
\end{equation}
Coefficients $u_n,v_n$  of the Bogolyubov transformation that appear in Eq.~(\ref{Lambda}) are given in turn by
\begin{equation}
u_n,v_n  = \sqrt{\frac{1}{2}\Bigl(\frac{A_n}{\omega_n}\pm 1\Bigr)}\,,\quad
A_{1,2}  =  S \Bigl[4J_\perp + 2E + (E-D\pm 4J_\perp)\sin^2\!\theta\Bigr]\,.
\end{equation}

\subsection{C. Fitting procedure and comparison with experiments}

Fits were performed by using a global fitting process with parameters sharing option (OriginPro 2021 software). We first performed ``individual'' fits, for each magnetic field configuration (along the a, b and c-axis), by considering  our complete data set and using the appropriate equations below and above the critical field to fit both magnon modes. During the initial fit run, the critical field position defining the range of applicability of each equations was first approximately defined from our experimental data. In the following iterations, for the fit with $B$ along the a and c-axis, all parameters were set free for adjustment, redefining the critical fields at each step, and eventually leading to a very good adjustment of each equation's range of validity around the critical fields. For the fit with $B$ along the b-axis, an upper value for the $J_\perp$ parameter (directly determining in this configuration the critical field) had to be imposed, based on the experimental $B_{\rm cr}^{(b)}$, to obtain a good fit (most likely because of the strong discontinuity of the magnon energy observed at the AFM-FM transition).

\begin{figure*}[h]
\begin{center}
\includegraphics[width=18cm]{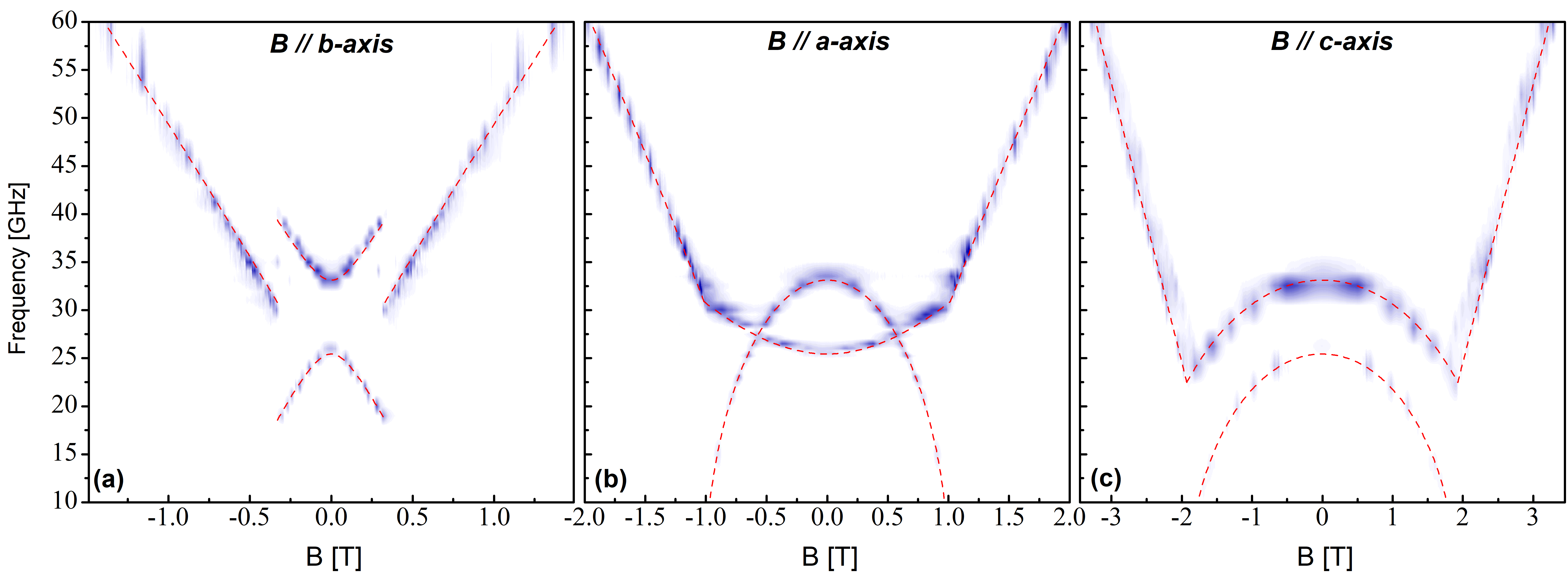}
\end{center}
\caption{(color online) Absorption spectra for a magnetic field B along the b (left panel), a (center panel) and c (right panel) axis. Color plot of the phase variation of the thermometer photo-resistance as a function of microwave frequency and magnetic field $B$, The phase difference with respect to its median value (which corresponds to the ``off-resonance'' value) is color-coded from zero (white) up to its measured maximum value (dark blue). Fits (red dashed-lines) obtained with our microscopic model and a \textit{single} set ($J_\perp$=0.069, $D=$0.396, and $E=$0.207) of ``global'' parameters (see text for details).}\label{globalfit}
\end{figure*}

The ``individual'' fit parameters are recalled in table \ref{table1}. ``Global'' fits were then performed to obtain a description of our entire data set (along the a, b and c-axis) with a \textbf{single} set of
parameters. Due to the enhanced constraints imposed on the parameters by the full data set, all parameters values could initially be set free, and converged to the values reported in the last line and the columns labelled as ``global'' in Table \ref{table1}. Fits obtained with these global parameters are reported together with experimental data (the same as figure 2 in the main text) in Fig.~\ref{globalfit}.
As can be seen, the fit of the full data set with a single set of parameters $J_\perp=$0.069, $D=$0.396, $E=$0.207 (and the g-factors respective to each axis) is very good, leading to low errors on the extracted fitting parameters (reported in table \ref{table1}). We also report in table \ref{table2} the corresponding values of interaction parameters for a different single-ion anisotropy Hamiltonian parametrization.
\begin{table}[h]
\begin{center}
\begin{tabular}{cccccccc}
\hline\hline $Axis$ & $J_\perp$(K) & $D$(K) & $E$(K)& $B_{\rm cr}$(T) & $B_{\rm cr}^{global}$(T) & $g_\alpha$ & $g_\alpha^{global}$ \\
\hline
b (easy) & 0.0734$\pm$0.0032 & 0.3891$\pm$0.009 & 0.2038$\pm$ 0.0118 & 0.346$\pm$0.015
 & 0.330$\pm$0.003 & 1.8969$\pm$0.0341 & 1.8778$\pm$0.0113   \\
a (intermediate)& 0.0691$\pm$0.0005  & 0.3993$\pm$0.0027  & 0.2151$\pm$0.0042 & 1.015$\pm$0.014 & 1.016$\pm$0.011 & 2.026$\pm$0.0134 & 2.0464$\pm$0.0107 \\
c (hard)& 0.0677$\pm$0.0008 & 0.387$\pm$0.0038 & 0.181$\pm$ 0.0064 & 1.902$\pm$0.018 & 1.927$\pm$0.013  & 1.9697$\pm$0.0174 & 2.0402$\pm$0.0097 \\
Global & 0.0694$\pm$0.0005 & 0.3956$\pm$0.0022 & 0.2074$\pm$0.0034 & - & - &- &- \\

\end{tabular}
\end{center}
\caption{CrSBr microscopic magnetic parameters and saturation fields for the three principal crystallographic directions (b-axis (easy), a-axis (intermediate), and c-axis (hard)).
Inter-layer exchange parameter $J_\perp$, anisotropy parameters $D$ and $E$, g-factor $g_\alpha$, and critical fields $B_{\rm cr}$. The values are extracted for each configuration at $T=5.4$~K. Parameters labeled as ``global'' are the ones  extracted from global fits of the magnetic field dependences along all three axes (see text). Parameter values and  and error bars are given here with the fit-determined accuracy.} \label{table1}
\end{table}
\begin{table}[h]
\begin{center}
\begin{tabular}{cccccc}
\hline\hline $J_\perp$(K) & $D$(K) & $E$(K)& $D_x$(K) & $D_y$(K)  & $D_z$(K) \\
\hline
0.0694 & 0.3956 & 0.2074 & -0.0755 & -0.2637 & 0.3393  \\

\end{tabular}
\end{center}
\caption{CrSBr ``global'' microscopic parameters for different single-ion anisotropy Hamiltonian parametrizations. Inter-layer exchange parameter $J_\perp$, anisotropy parameters $D$ and $E$ (as defined in the main text) and corresponding values for the parameters $D_{x}=-E+D/3$, $D_{y}=-2D/3$, and $D_{z}=E+D/3$ associated with the Hamiltonian $\hat{\mathcal{H}}_{\rm SI} \simeq [D_xS^{x2}+ D_yS^{y2}+D_zS^{z2}]$.} \label{table2}
\end{table}
We note that since $J_\perp$ is significantly smaller than the single ion anisotropy, the transition temperature in CrSBr could be determined by \textit{anisotropy} together with the in-plane exchange. Further theoretical studies will be necessary to substantiate this proposal.

Let's finally comment on how our model can describe other recently reported magnon dispersion data \cite{Cham2022,Bae2022}. The low magnetic field a-axis and b-axis data reported in Ref.~\onlinecite{Cham2022} can be nicely reproduced with our model, which provides the same functional magnetic field dependence as the formalism used in this work. The c-axis data presented in Ref.~\onlinecite{Bae2022} can be fitted by our model, with a less good agreement for the low field part of the upper energy branch if the g-factor is set to stay close to 2.

When letting all parameters initially free, the values of $E$ extracted from these fits is significantly higher than the one of our experiments. $D$ is also higher, but to a lower extent, and $J_\perp$ tends to be slightly higher. As a consequence, the condition $D-E> 2J_\perp$ is not as clearly satisfied as in our case, implying a possibly higher proximity to the spin-flop phase. Nevertheless, the g-factor extracted from this process
are far apart from 2, and most likely, due to the absence of data in the saturated regime (where g-factors play a key role), not reliable. We therefore repeated the same process, but by fixing this time the g-factor values
to the ones determined at higher magnetic fields in our experiments. In this case, the extracted $J_\perp$ are significantly stronger than in our experiment ($J_\perp$ closer to 0.09 K), while $D$ is slightly smaller and $E$ only slightly smaller. The consequence on the $D-E$ vs $2J_\perp$ comparison is the same: the two quantities are closer than in our experiments, implying a higher proximity to the spin-flop phase. Even though the data from Ref.~\onlinecite{Cham2022,Bae2022} were collected in a restricted magnetic field range compared to our experiment (which for example precludes a reliable estimation of the g-factors), we believe there is a small intrinsic difference between the type of samples studied, since restricting the fitting range to lower magnetic fields in our case does not change dramatically the parameters hierarchy. Such a difference could emerge from different growth conditions, and/or from the existence of local structural inhomogeneities, as it has been observed that the exchange parameter $J_\perp$ in CrSBr is sensitive to local strains~\cite{Cenker2022}.

\end{document}